# Organometallic Complexes of Graphene

Santanu Sarkar, Sandip Niyogi, Elena Bekyarova and Robert C. Haddon*

*Center for Nanoscale Science and Engineering, Departments of Chemistry and Chemical & Environmental Engineering, University of California, Riverside, CA 92521-0403 (USA)*
E-mail: haddon@ucr.edu

**Abstract:** We demonstrate the organometallic hexahapto ($\eta^6-$) complexation of chromium with graphene, graphite and carbon nanotubes. All of these extended periodic π-electron systems exhibit some degree of reactivity toward the reagents $Cr(CO)_6$ and ($\eta^6-$benzene)$Cr(CO)_3$, and we are able to demonstrate the formation of ($\eta^6-$arene)$Cr(CO)_3$ or ($\eta^6-$arene)$_2$Cr, where arene = single-walled carbon nanotubes (SWNT), exfoliated graphene (XG), epitaxial graphene (EG) and highly-oriented pyrolytic graphite (HOPG). We find that the SWNTs are the least reactive presumably as a result of the effect of curvature on the formation of the hexahapto bond; in the case of HOPG, ($\eta^6-$HOPG)$Cr(CO)_3$ was isolated while the exfoliated graphene samples were found to give both ($\eta^6-$graphene)$_2$Cr, and ($\eta^6-$graphene)$Cr(CO)_3$ structures. We report simple and efficient routes for the mild decomplexation of the graphene−chromium complexes which appears to restore the original pristine graphene state. This study represents the first example of the use of graphene as a ligand and is expected to expand the scope of graphene chemistry in connection with the application of this material in organometallic catalysis.

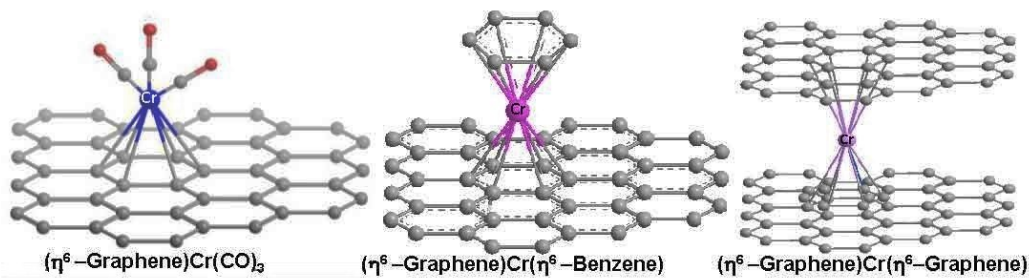

($\eta^6-$Graphene)$Cr(CO)_3$     ($\eta^6-$Graphene)$Cr(\eta^6-$Benzene)     ($\eta^6-$Graphene)$Cr(\eta^6-$Graphene)

**Introduction**

Graphene, a two-dimensional crystal of atomic thickness, has attracted significant attention in recent years because of its intriguing electronic structure, which provides the basis for unique physical phenomena.[1-4] The electronic properties of graphene are complemented by its mechanical and thermal properties and this has served to further raise interest in the material. In order to realize the potential applications of graphene, there is a need to develop the chemistry and to tailor its physical properties.

Apart from noncovalent functionalization reactions, recent reports on the chemistry of graphene include Diels-Alder reactions,[5] and covalent bond forming reactions with aryl groups,[6-8] which occur under very mild conditions and may provide either a monolayer coverage,[6,9] or multilayer structures of various thicknesses.[10] The traditional derivatives, graphene oxide,[11-20] graphane (completely hydrogenated graphene)[21,22] and fluorographene $((CF)_n)$[23-27] have also been studied. In these reactions the $sp^2$ hybridized carbon atoms of graphene are converted to $sp^3$ hybridization; breaking the π-conjugation length in this manner is a route for band gap engineering of graphene.[9,28] The flat π-surface of graphene together with its unique Fermi level electronic structure suggests the exploration of chemical reactions involving other bonding configurations. In hexahapto-metal complexation reactions, the vacant d-orbitals of a transition metal can overlap with and accept electron density from the occupied π-orbitals of graphene, resulting in the coordination of the metal atom to individual benzenoid aromatic rings of graphene and in bis-graphene metal sandwich complexes it is possible to electronically conjugate adjacent graphene sheets. The $\eta^6$–metal bond to graphene is not expected to significantly rehybridize the $sp^2$ carbon atoms and can be anticipated to provide a much milder modification of the electronic structure of graphene than σ-bond formation; as a result, the metal coordination site on the graphene surface is expected to be mobile.

In prior work, $\eta^2$–complexation of single-walled carbon nanotubes (SWNTs) was explored in reactions with Vaska's complex $[Ir(CO)Cl(PPh_3)_2]$[29,30] and Wilkinson's complex.$[RhCl(PPh_3)_2]$[30,31] The Rh-based SWNT-Wilkinson's complex showed recyclable catalytic hydrogenation of cyclohexene at room temperature.[31] Chromium complexes of HiPCO single−walled carbon nanotubes have also been discussed.[32] In general, the curvature of the conjugated carbon network enhances the reactivity of the fullerenes and carbon nanotubes (where addition chemistry predominates),[33, 34] including the dihapto-metal complexation reactions to give coordination compounds such as $\eta^2-\pi$ complexes of $C_{60}$, namely $\eta^2-C_{60}$, $\mu-\eta^2:\eta^2-C_{60}$, $\mu_3-\eta^2:\eta^2:\eta^2-C_{60}$ and $\eta^2$ and $\eta^1$ $\sigma-\pi$ mixed complexes, namely $\mu_3-\eta^1:\eta^2:\eta^1-C_{60}$, $\mu_3-\eta^1:\eta^1:\eta^2-C_{60}$, together with similar complexes of higher fullerenes.[35] However, it has been shown that the curvature in $C_{60}$ significantly inhibits the potential of the molecule to function as a ligand in pentahapto- ($\eta^5$), and hexahapto- ($\eta^6$) complexation reactions because the fullerene $\pi$-orbitals are directed away from the metal as a result of the rehybridization of the ring carbon atoms.[36, 37] Nevertheless, the curvature can be ameliorated by functionalization,[37-39] and this has allowed the preparation of organometallic fullerene derivatives.[38, 40] In $C_{60}$ the $\pi$-orbital axis vectors are directed away from the center of the respective rings and make angles of 16.3° (POAV2) to a normal to the plane of the five-membered rings and 25.8° (POAV2) to a normal to the plane of the six-membered rings; hence, hexahapto- ($\eta^6$) coordination is even more strongly disfavored than pentahapto- ($\eta^5$) complexation.[37] However, an examination of carbon nanotube structures, which have lower curvature than the fullerenes (carbon nanotubes are curved in 1-dimension whereas fullerenes are curved in 2-dimensions),[33, 34] together with the previous report on the complexation of HiPCO SWNTs encouraged us to explore the organometallic reactions of electric arc (EA)-SWNTs, which have significantly larger diameter.[41, 42]

We begin by discussing the results of the reaction of EA-SWNTs with ($\eta^6$–benzene)Cr(CO)$_3$ and then move to an investigation of the reaction of various forms of graphite and graphene with Cr(CO)$_6$ and ($\eta^6$–benzene)Cr(CO)$_3$. These two reagents were chosen to explore the fundamental arene-metal complexation mechanisms,[43] which depend on the displacement of three carbon monoxide molecules and in the present manuscript we demonstrate that all of these forms of conjugated carbon exhibit some degree of reactivity, as delineated in Scheme 1. The study demonstrates the wide applicability of this type of chemistry in engineering the electronic structure of extended, periodic $\pi$-electron systems as well as expanding the scope of graphene chemistry towards applications in organometallic catalysis and electronics. Additionally, we demonstrate a number of routes for de-complexation that appear to restore graphene to its pristine state.

**Experimental Section**

**Chemicals and Instrumentation**

$Cr(CO)_6$ (98%, Aldrich, FW = 220.06, m.p. > 150 °C), ($\eta^6$–benzene)$Cr(CO)_3$ (98%, FW = 214.4, m.p. = 163-166 °C), ortho-dichlorobenzene (anhydrous, 99%), *p*-xylene (anhydrous, 99%), mesitylene (98%), microcrystalline graphite (1-2 micron, synthetic), dibutyl ether (anhydrous, 99.3%) were purchased from Sigma-Aldrich. HOPG was obtained from Union Carbide Corporation. The single-walled nanotube (SWNT) samples were obtained from Carbon Solutions Inc. (P2-SWNT, http://www.carbonsolution.com). The epitaxial graphene films grown on the C-face of SiC substrates (Cree Inc, 4H SiC) with dimensions of 3.5 × 4.5 $mm^2$ were provided by Professor Walt de Heer.[44, 45]

The ATR-IR spectra were taken using a Thermo Nicolet Nexus 670 FTIR instrument, equipped with an ATR sampling accessory. Raman spectra were acquired in a Nicolet Almega XR Dispersive Raman microscope with a 0.6 μm spot size and 532 nm excitation. STM images were acquired under ambient conditions using Pt/Ir tips in a DI Nanoscope IIIa multimode SPM. TGA was performed at a heating rate of 5 °C/min in air using a Pyris 1 thermo-gravimetric analyzer (Perkin Elmer). NMR spectra were collected in a Bruker 300MHz instrument and UV-vis spectra were collected in a Varian Cary 5000 spectrometer.

**Resistivity measurements**

The resistance was measured at room temperature in air with a Keithley 236 source-measure unit. The temperature dependence of resistivity was measured in a custom-made helium variable-temperature probe using a Lake Shore 340 temperature controller driven with custom LabVIEW software. The device was degassed for 12 hours in vacuum (2 x $10^{-8}$ torr) prior to the measurement.

**XPS Measurement**

The XPS measurements were performed with a Kratos Axis Ultra spectrometer (Kratos Analytical, Manchester, UK) using Al Kα monochromated radiation at 1486 eV at base pressure of about 3 x$10^{-8}$ Torr. The survey spectra were recorded using 270 watts of X-ray power, 80 pass energy, 0.5 eV step size. A low-energy electron flood from a filament was used for charge neutralization. The spectra are shown without energy-scale correction.

**Preparation of the chromium–SWNT complex, ($\eta^6$–benzene)Cr($\eta^6$–SWNT)**

($\eta^6$–benzene)Cr(CO)$_3$ (17 mg, 0.08 mmol, FW = 214.4) was added to a suspension of purified SWNTs (18 mg, 1.5 mmol of carbon atoms; purified SWNTs – P2-SWNT) in dry distilled THF (6 mL). The reaction mixture was sonicated for 2 min using an ultrasonic probe (Cole-Parmer, 50% amplitude) and then degassed with argon for 15 min in absence of light. The reaction mixture was heated at 72 °C for 72 h in the dark under a positive pressure of argon, after which it was cooled to room temperature. The suspension was filtered through a 0.2 µm PTFE membrane and the solid was washed with anhydrous ether. The resulting chromium-SWNTs complex (~21 mg, isolated yield) was dried overnight under vacuum in the dark.

**Exfoliation of microcrystalline graphite**

Microcrystalline graphite (1-2 µm, 500 mg, synthetic, Sigma-Aldrich) was sonicated in o-dichlorobenzene (~200 mL ODCB) for 1 h using a probe ultrasonic processor (Cole-Parmer) at 40% amplitude. The dispersion was centrifuged at 14000g for 30 min. The resulting supernatant (which yielded dispersions of graphene in o-dichlorobenzene) [46] was collected and concentrated under vacuum. The powdered exfoliated graphene was dried in high vacuum overnight and used for subsequent reactions after re-dispersion in dry, distilled THF.

**Reaction of exfoliated graphene and ($\eta^6$–benzene)Cr(CO)$_3$**

In a typical reaction, ($\eta^6$–benzene)Cr(CO)$_3$ (36 mg, 0.17 mmol, FW = 214.4) was added to a suspension of exfoliated graphene (20 mg, 1.67 mmol of carbon) in THF (4 mL). The reaction mixture was stirred vigorously and refluxed at 72°C under argon, in the absence of light for 48 h. The resulting mixture was filtered using 0.2 µm PTFE filter paper and the solid was washed with fresh THF and ether (to remove excess chromium reagent). The resulting solid was dried under vacuum overnight in dark to obtain a silver-colored solid (~27 mg of solid was isolated).

**Reaction of exfoliated graphene and Cr(CO)$_6$**

Cr(CO)$_6$ [8.2 (47.8) mg, 0.04 (0.22) mmol, FW = 220.06] was added to a suspension of exfoliated graphene (20 mg) in THF [4 (10) mL] and dibutyl ether [2 (5) mL]. The black suspension was stirred vigorously and refluxed at 72 °C in the absence of light under an atmosphere of argon for 48 h. The reaction mixture was filtered using 0.2 μm PTFE filter paper and the solid was washed with fresh THF and ether. The resulting solid [~19 mg] was dried under vacuum overnight in the dark.

**Reaction of HOPG and EG with $Cr(CO)_6$**

HOPG (~0.28 $cm^2$) or EG on 4H-SiC (3.5mm x 4.5mm), was heated in a solution of $Cr(CO)_6$ (30 mg, 0.14 mmol) in THF (3 mL) and dibutyl ether (1 mL) under a positive pressure of argon at 72°C for 48 h without stirring, then washed with anhydrous ether and dried under a gentle flow of argon.

**Reaction of HOPG and EG with ($\eta^6$–benzene)$Cr(CO)_3$**

A piece of HOPG (~0.32 $cm^2$) or EG on 4H-SiC (3.5mm x 4.5mm) was heated in a solution of ($\eta^6$–benzene)$Cr(CO)_3$ (33 mg, 0.16 mmol) in THF (3 mL) under a positive pressure of argon at 72°C for 72 h without stirring, after which the sample was washed with THF and anhydrous ether and dried under a gentle flow of argon.

**De-complexation of graphene−Cr complexes by ambient oxidation**

To collect the Raman spectra of the graphene−Cr complexes, a dispersion of the sample was allowed to dry on a $SiO_2$ substrate; the color contrast with the substrate allowed identification of graphene samples of various thickness. After recording the Raman spectra of the products, the decomplexation reaction was carried out by adding a few drops of acetonitrile to the substrates and exposing them to sunlight, under a glass petridish; Raman spectroscopy was used to follow the progress of the de-complexation reaction.

**De-complexation with electron rich arenes**

The Cr complexes of XG and HOPG were either refluxed or warmed (oil bath temperature of 100 °C for benzene, 150 °C for *p*-xylene and 150 °C for mesitylene) with the arene (~5 mL) under argon overnight. The resulting reaction mixture was filtered through a 0.2 μm PTFE membrane and the solid was washed with a copious amount of anhydrous diethylether. The resulting solid was dried for 1 h under vacuum and characterized by Raman spectroscopy.

**Results and Discussion:** In metal carbonyl complexes and their ligand exchanged derivatives, the frequency of the C–O stretching vibration can be used to distinguish between terminal ($\nu_{CO}$ = 2000 ± 100 cm$^{-1}$) and bridged ($\nu_{CO}$ = 1800 ± 75 cm$^{-1}$) CO structures as well as providing information on the electronic environment of the metal center.[47, 48] In SWNT, HOPG and graphene chemistry the use of IR spectroscopy is sometimes limited by weak spectral features as a result of the dynamic dipole moments and IR assignments of even well documented complexes can be difficult.[49] In the case of metal complexes of polyaromatic arenes, haptotropic slippage[50] and fluxional behavior[51] of the ligands further complicates characterization using FT-IR spectroscopy.[43] Scanning tunneling microscopy provides more direct evidence of chemical functionalization, if well ordered samples from optimized reactions are available.[52-54] We have made use of FT-IR spectroscopy in characterizing the products of the reaction of SWNTs, HOPG, and graphene with organometallic precursors, together with Raman microscopy (excitation at λ = 532 nm) and absorbance spectroscopy which are well known in carbon materials science. In SWNTs, the inter-band transitions in metallic and semiconducting structures give rise to distinct peaks in the UV-vis-NIR absorbance spectrum and modification of the π-conjugation affects these band transitions. For example, in the side-wall functionalization of SWNTs with dichlorocarbene, the change in the absorbance spectrum was related to the degree of functionalization, showing that after complete reaction all band transitions are lost.[55] Similar effects have been demonstrated by other groups using different chemical reactions.[56] In graphitic materials, the frequency, width and spectral shape of the most intense Raman peaks, labeled as the G and 2D bands, can be used to determine the number of layers, their inter-layer electronic interaction as well as the Fermi-level electronic structure due to the resonance enhancement of these vibrations.[57] In graphene samples prepared using various techniques, the intensity of the D-band has been found to be a useful signature for following the effect of covalent chemical reactions.[7-9, 21, 58] The D-band is due to the in-plane breathing vibration of conjugated six-membered rings and this feature is not observed without some disruption of the long range periodicity of the graphene lattice; in covalently functionalized graphene, an empirical relationship between the density of sp$^3$ centers and the ratio of the intensities of the G and D bands, has been reported.[9, 59, 60] The detailed understanding of the D-band in terms of molecular vibrations in various bonding environments is also of interest in small molecule chemistry,[61-63] and the present study clearly relates to this work.

**Scheme 1** Representative reactions of SWNT, graphite and graphene with ($\eta^6$-benzene)Cr(CO)$_3$ and Cr(CO)$_6$ based on product structures.

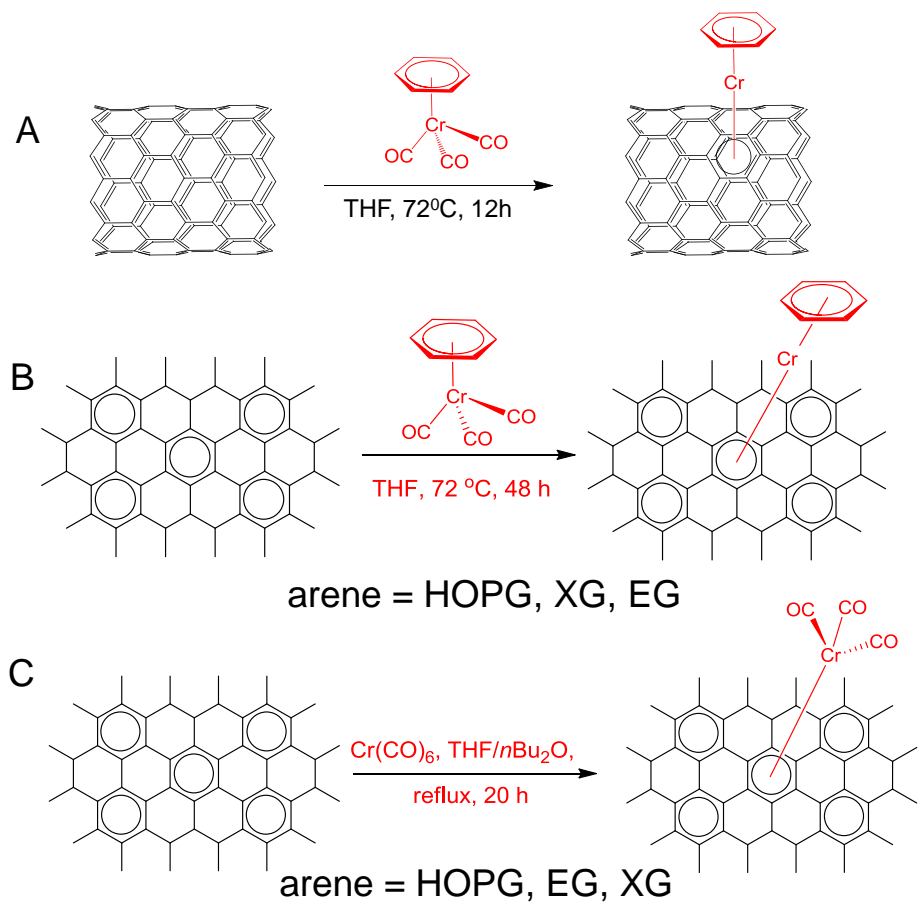

### Reaction of ($\eta^6$−benzene)Cr(CO)$_3$ with EA-SWNTs

The reaction of EA-SWNTs (average diameter $D_{av}$ = 1.55 ±0.1 nm),[41, 42] with ($\eta^6$−benzene)Cr(CO)$_3$ in tetrahydrofuran (THF), which is illustrated in Scheme 1A, gave rise to a black powder that was isolated by filtration.

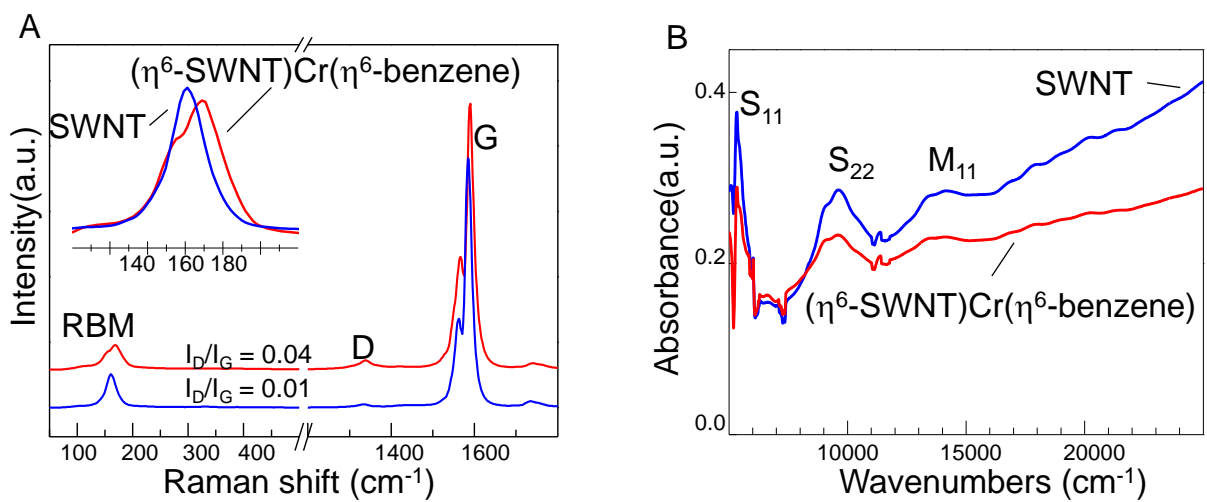

**Fig. 1.** Characteristics of the ($\eta^6$–SWNT)Cr($\eta^6$–benzene) complex of single−walled carbon nanotubes. A) Raman spectra of the starting SWNTs and the products, collected with $\lambda_{EX}$ = 532 nm on solid samples. The inset shows the RBM region of the spectra; B) Absorbance spectra of the starting SWNTs and the products, collected on dispersions in dimethylformamide; the dispersions were prepared at approximately similar optical densities and the spectra are not normalized. The features on the lower energy side of the $S_{11}$ band and on both sides of the $S_{22}$ bands are due to water in the solvent.

The changes in the Raman spectrum of SWNTs due to reaction is shown in Fig. 1A; the intensity of D-band increases relative to the G-band as previously observed for a sidewall functionalization process;[55] ($I_D/I_G$ ~0.04 as compared to $I_D/I_G$ ~ 0.01 in the pristine SWNTs). The SWNT radial breathing mode (RBM) is resonantly enhanced by interband electronic transitions and the frequency is inversely proportional to the diameter. Thus, when the SWNTs are chemically functionalized the band transition energies are modified and this may affect the resonance conditions in the Raman experiment and in cases where the chemical reaction is dependent on nanotube diameter and chirality the RBM band profile takes on a different shape due to changes in the resonance conditions of the various SWNT populations.[64] The inset in Fig. 1A shows such a change in the RBM profile; although nanotube chiralities cannot be assigned from a single excitation Raman spectrum, the loss of intensity at the lower frequency of the RBM band indicates that the larger diameter SWNTs are preferentially removed from resonance by the chemical reaction.

The UV-vis-NIR absorption spectrum of the reaction product (Fig. 1B) shows a decrease in the intensities of all interband transitions; this is most clearly seen for the second semiconducting interband transition ($S_{22}$). It is apparent that the intensities of the larger diameter SWNTs are preferentially weakened in the product in accord with the previous discussion, which suggests that lower curvature structures will be the most reactive. The change of the SWNT spectra on reaction with benzene chromium tricarbonyl is qualitatively similar to that observed on side-wall functionalization with dichlorocarbene,[55] although the reaction does not proceed to the same degree, perhaps due to the incomplete dispersal of the current sample. The ATR-IR spectrum of the product (Supporting Information, Fig. S1) does not show the C–O vibrations, but it will be of some interest to determine the mode of complexation as it is possible that chromium could bind to the interior wall of the carbon nanotubes.[37]

In order to examine the effect of the complexation of SWNTs with Cr on the electronic structure of the material, free-standing films of pristine SWNTs and the ($\eta^6$–SWNT)Cr($\eta^6$– benzene) product were prepared and transferred to a glass substrate with pre-deposited gold contacts.[65] The SWNT film thickness was estimated from the near-IR spectra of the films (absorbance at 550 nm)[66-68] The conductivity of the functionalized SWNTs ($\sigma_{RT} \sim 100$ S cm$^{-1}$) decreased by a factor of 3 from the pristine value of $\sigma_{RT} \sim 300$ S cm$^{-1}$.[68]

### Reaction of Cr(CO)$_6$ and ($\eta^6$– benzene)Cr(CO)$_3$ with exfoliated graphene (XG)

The exfoliated graphene (XG) sample consists of a mixture of multilayer-graphene flakes (micrometer dimensions), together with single layer graphene, as evidenced by comparing the intensities of the 2D and G bands in the spectra shown in Fig. 2A and 2B and the lineshape of the 2D band.

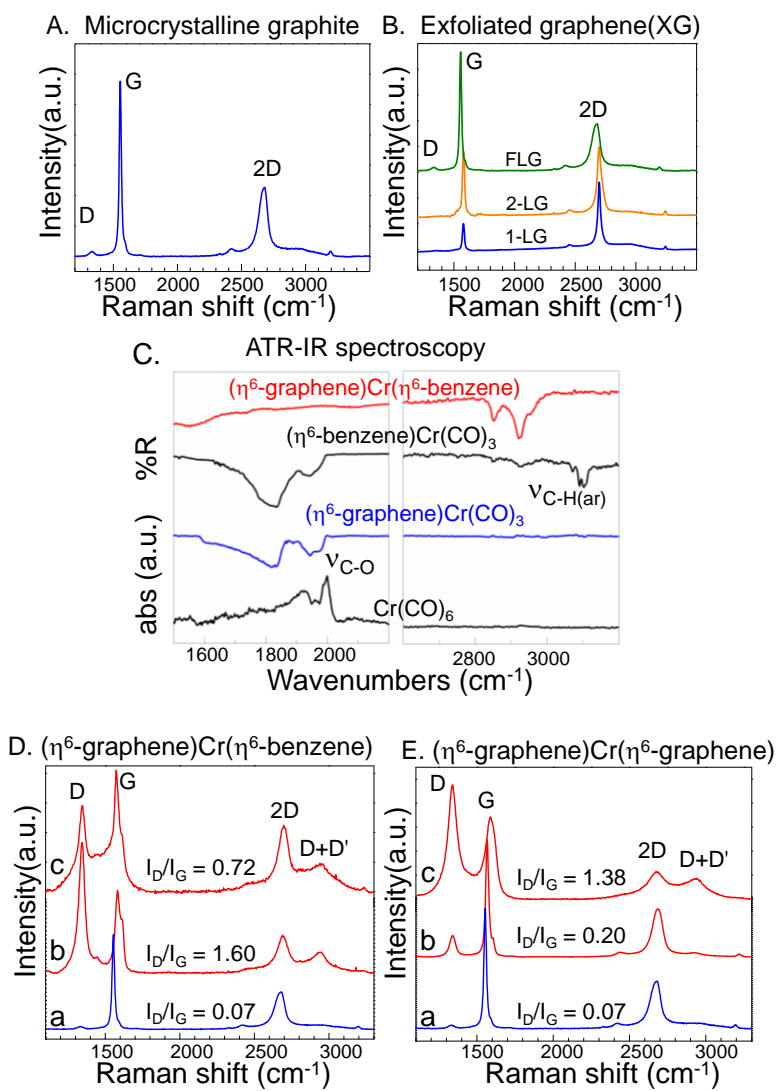

**Fig. 2.** Characteristics of the exfoliated graphene samples prepared by sonication and centrifugation of microcrystalline graphite in o-dichlorobenzene and complexes. A) Raman spectrum of the starting microcrystalline graphite. B) Raman spectra of the sample after sonication in ODCB and centrifugation, showing characteristics of few layer graphene (FLG), bilayer graphene (2-LG) and single layer graphene(1-LG). C) FT-IR spectra of the starting chromium reagents and the ($\eta^6$–graphene)Cr($\eta^6$–benzene) product. D) Raman spectra of the ($\eta^6$-graphene)Cr($\eta^6$–benzene) product, collected on solid samples; spectra collected from different graphene flakes (spectra b and c) are shown together with that of the starting material (spectrum a); E) Raman spectra of ($\eta^6$–graphene)Cr($\eta^6$–graphene) product collected on solid samples; spectra collected from different regions (spectra b and c) of the sample are shown together with that of the starting material (spectrum a).

After reaction of XG with 0.13 equivalents of $Cr(CO)_6$ (Scheme 1C) the IR spectrum showed C–O stretching vibrations at 1939 cm$^{-1}$ (Fig. 2C and Supporting Information Fig. S2), compared to those in $Cr(CO)_6$ at 2000 cm$^{-1}$. The $Cr_{2p}$ XPS spectra show the presence of Cr(0) (Supporting

Information Fig. S3) and the product is assigned as ($\eta^6$–graphene)Cr(CO)$_3$. After the reaction of XG with 0.02 equivalents of Cr(CO)$_6$, the C–O vibrations could not be observed in the product (Supporting Information, Fig. S2) but the Cr$_{2p}$ XPS spectra shows the presence of Cr(0) (Supporting Information Fig. S3) and we assign the structure of the product as ($\eta^6$–graphene)Cr($\eta^6$–graphene). These observations of varying product structures as a function of the reagent stoichiometry is consistent with those reported for molecular chromium complexes.[69, 70] The ATR-IR spectrum (Supporting Information Fig. S4) of the reaction product of XG and ($\eta^6$–benzene)Cr(CO)$_3$ showed the aromatic C–H vibration of benzene at lower frequency compared to that in the starting material[71] and does not show the C–O vibrations, indicating the formation of the ($\eta^6$–graphene)Cr($\eta^6$–benzene) product (Fig. 2C).

In Figs 2D and 2E, the changes in the Raman spectra of XG after reaction with ($\eta^6$–benzene)Cr(CO)$_3$ and Cr(CO)$_6$ are compared. Both reactions show a strong D-band accompanied by (D+D') peaks. The appearance of strong D peaks in the Raman spectra of the products is attributed to the chromium complexation of graphene via $\eta^6$-bond formation and associated changes in the electronic structure of graphene. The spectra signify an entirely distinct electronic structure from that of nitrophenyl-functionalized graphene, which gives sharp Raman peaks even at high coverage densities.[9]

Thermogravimetric analysis of ($\eta^6$–graphene)Cr($\eta^6$–benzene) in air left a residue of 1.8 wt% (Supporting Information, Fig. S5), which we attribute to the presence of chromium metal that is converted to Cr$_2$O$_3$ (recovered as a dark green residue) during pyrolysis and corresponds to approximately 0.2 atomic % Cr or 1 Cr atom per 500 C atoms. The low yield (metal loading) of the reaction may be due to the presence of incompletely exfoliated graphite nanoparticles;[72, 73] in the case of multilayer graphene particles only the outermost surfaces can participate in the complexation reaction. Thus, XPS analysis showed presence of 1.3 atomic % Cr in this material.

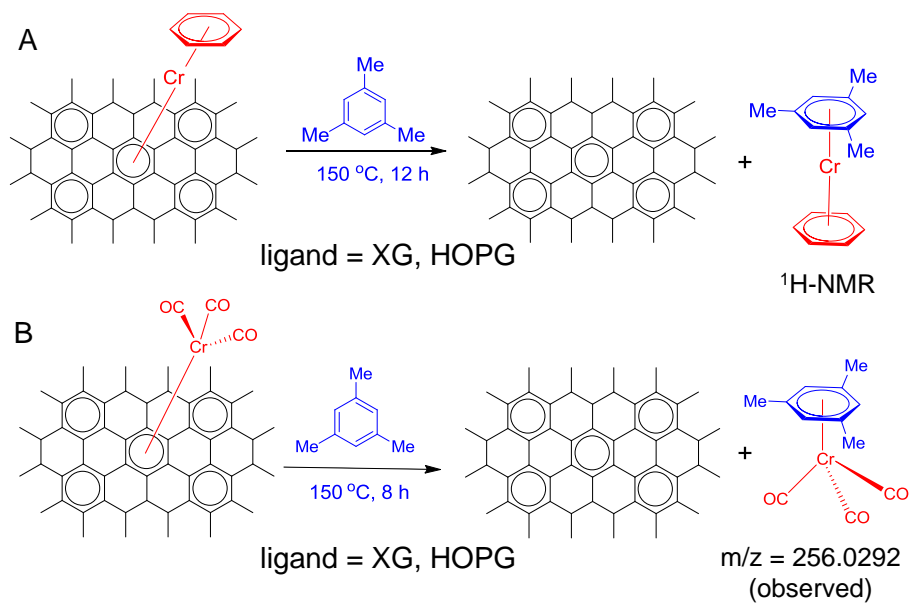

**Scheme 2.** Decomplexation of graphene–chromium complexes by substitution reactions with mesitylene.

# Regeneration of exfoliated graphene from ($\eta^6$–graphene)Cr(CO)$_3$ and ($\eta^6$–graphene)Cr($\eta^6$– benzene)

The organometallic chemistry of (arene)metal(CO)$_3$ complexes provides a number of routes for the one-step derivatization of arene molecules, which is of particular interest in the development of graphene chemistry.[43, 74] The simplest of these are the arene decomplexation or arene exchange reactions (Scheme 2); the arene–metal bond in (arene)Cr(CO)$_3$ complexes is cleaved upon oxidation of the metal by exposure of the complex to sunlight in air in the presence of diethyl ether or acetonitrile, to yield the free arene.[75] Refluxing (arene)Cr(CO)$_3$ complexes in the presence of electron-rich arenes such as mesitylene and pyridine, can also yield arene exchanged Cr(CO)$_3$ complexes.[75] These decomplexation reactions are of interest because they provide a route to reverse the changes in the graphene electronic properties.

Fig. 3A shows that the ratio of the intensities of the D and G bands ($I_D/I_G$) in the spectra of the ($\eta^6$–graphene)Cr($\eta^6$–graphene) complex decreases to 0.06 after exposure to light for 3 hours in acetonitrile (compared to $I_D/I_G$ = 1.59 for the same complex before exposure to sunlight). The de-complexation of ($\eta^6$–graphene)Cr($\eta^6$–benzene) requires longer exposure to sunlight; after exposure for 6 hours the $I_D/I_G$ ratio decreases from 1.40 to 0.18 (Fig. 3B) suggesting that the ($\eta^6$–graphene)Cr($\eta^6$–graphene) complex is more reactive than ($\eta^6$–graphene)Cr($\eta^6$–benzene). The strength of the ($\eta^6$–graphene)–Cr bond was investigated in a series of competition reactions with electron rich arenes.[76] Refluxing the ($\eta^6$–graphene)Cr($\eta^6$–benzene) complex ($I_D/I_G$ ~ 1.1, Fig. 3C) in benzene (oil bath temperature 100 °C; final $I_D/I_G$ = 0.85) or *p*-xylene (oil bath temperature 150 °C; final $I_D/I_G$ = 0.45) was insufficient to fully regenerate the XG, whereas heating in mesitylene (oil bath temperature 150 °C, Scheme 2A)  gave a final $I_D/I_G$ = 0.04, indistinguishable from the starting graphene sample.  Thus, it is apparent that the ($\eta^6$–graphene)–Cr bond is fairly robust; the ($\eta^6$–benzene)–Cr bond energy is reported to be 164 kJ mol$^{-1}$.[76]

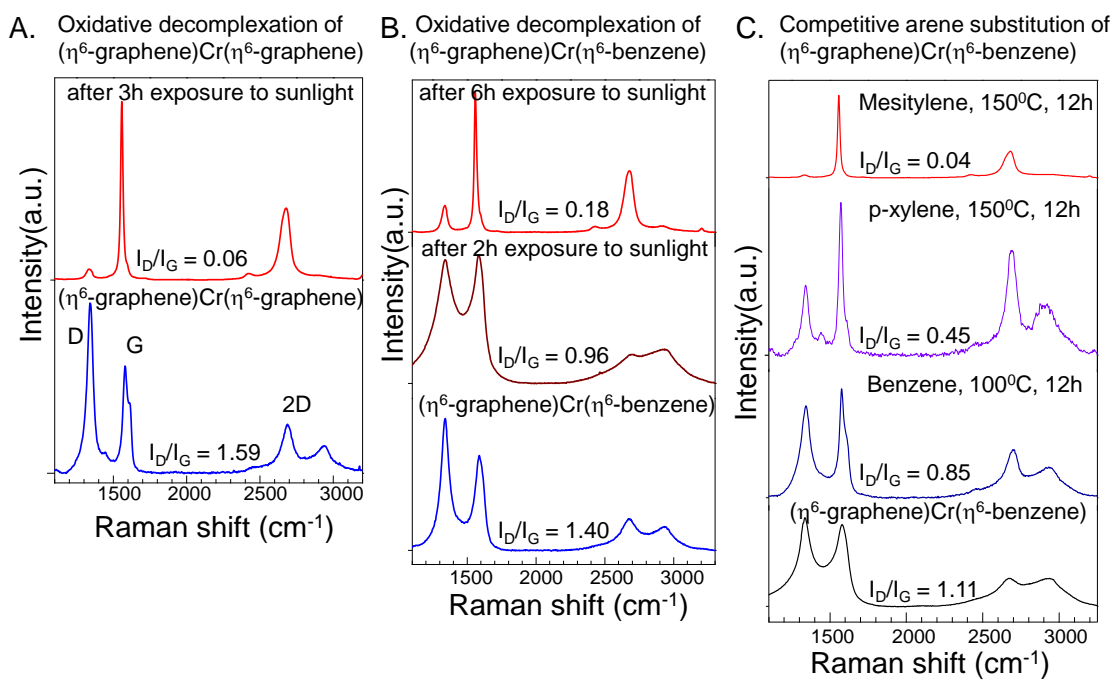

**Fig. 3.** Raman characteristics of the de-complexation reactions of exfoliated graphene (EG); changes of the ($\eta^6$–graphene)Cr($\eta^6$–graphene) product  (A) Effect of sunlight on ($\eta^6$–graphene)Cr($\eta^6$–graphene)  (B) Effect of sunlight on the ($\eta^6$–graphene)Cr($\eta^6$–benzene); (C) De-complexation of ($\eta^6$–graphene)Cr($\eta^6$–benzene) with benzene, p-xylene and mesitylene.

The product from the decomplexation reaction of ($\eta^6$–graphene)Cr($\eta^6$–benzene) in mesitylene was yellow in color, turned green on standing in air and eventually yielded a dark green precipitate (Supporting Information, Fig. S6) . The yellow extract in mesitylene was characterized using $^1$H NMR spectroscopy (Supporting Information Fig S7); based on the changes in the chemcial shifts of the aromatic protons and their integration values, the decomplexed product is assigned to ($\eta^6$–mesitylene)Cr($\eta^6$–benzene). The product was also analyzed using UV-vis absorption spectroscopy (Supporting Information Fig. S8) which showed charge transfer bands between arene and chromium in the yellow solution and FAB-MS showing a m/z = 282.2862 (Supporting Information, Fig. S9).

**Reaction of Cr(CO)$_6$ and ($\eta^6$–benzene)Cr(CO)$_3$ with epitaxial graphene (EG) and highly oriented pyrolytic graphite (HOPG)**

Compared to the high surface area exfoliated graphene (XG), HOPG and epitaxial graphene are millimeter scale samples with long range crystalline order and well defined surfaces; their distinguishing feature is the interlayer stacking pattern. In HOPG, the graphene layers are Bernal stacked with strong interlayer electronic interaction, resulting in the formation of two triangular sublattices in the AB plane as seen in the STM image in Fig. 4A. In epitaxial graphene grown on SiC wafers, the graphene layers are rotationally disordered, eliminating the periodic electronic interaction in the c-axis direction and the STM image of the surface shows the typical √3 x √3 reorganization (Fig. 4B). Beyond the development of chemistry to modify the physical properties of graphene, it is interesting to compare the chemical consequences of interlayer stacking in epitaxial graphene and graphite.[8]

The IR C–O stretch in ($\eta^6$–HOPG)Cr(CO)$_3$ was observed at 1948 cm$^{-1}$ while in the starting Cr(CO)$_6$ complex it was observed at 2000 cm$^{-1}$ and in ($\eta^6$–benzene)Cr(CO)$_3$ the C–O stretching was observed at 1942 cm$^{-1}$ (Fig. 4C and Fig. S10 in Supporting Information). It is known that, when arene ligands substitute for CO in Cr complexes, the electron density on Cr increases and the CO vibration frequency decreases.[47] After reaction with ($\eta^6$–benzene)Cr(CO)$_3$ the C–O vibrations could not be identified and the aromatic C–H vibrations are observed at lower frequency than that of the reactant (Fig. 4D and Supporting Information, Fig. S10).[71]
In contrast to the chromium complexes of exfoliated graphene, the Raman spectrum of the ($\eta^6$–HOPG)Cr(CO)$_3$ and ($\eta^6$–HOPG)Cr($\eta^6$–benzene) showed a weak D-band with $I_D/I_G$ = 0.18 and 0.10 respectively (Fig. 4E); treatment of EG with $Cr(CO)_6$ also gave rise to very weak D-bands with $I_D/I_G = 0.10$ (Fig. 4F). The reaction of $Cr(CO)_6$ with a pre-wired EG sample led to change in the transport properties of epitaxial graphene; the four point resistance of the EG sample increased by about 30% after treatment with $Cr(CO)_6$ (Fig. 5G), as compared to the factor of two increase in resistance that was observed on nitrophenyl functionalization.[6] The EG samples on SiC wafers do not provide a compatible configuration for ATR-IR measurements, so X-ray photoelectron spectroscopy (XPS) was used to characterize the chemical composition of the products. As shown in Fig. 4H, the survey scan of pristine EG reveals a strong C-peak at 284.5 eV corresponding to $sp^2$ carbon and a weak O1s peak presumably due to oxygen atoms bound to carbon at the graphene edges. The spectra of $(\eta^6–EG)Cr(\eta^6–benzene)$ and $(\eta^6–EG)Cr(CO)_3$ show distinct peaks at ~576.7 eV and ~586.3 eV, which are assigned to Cr $2p_{3/2}$ and Cr $2p_{1/2}$, respectively.

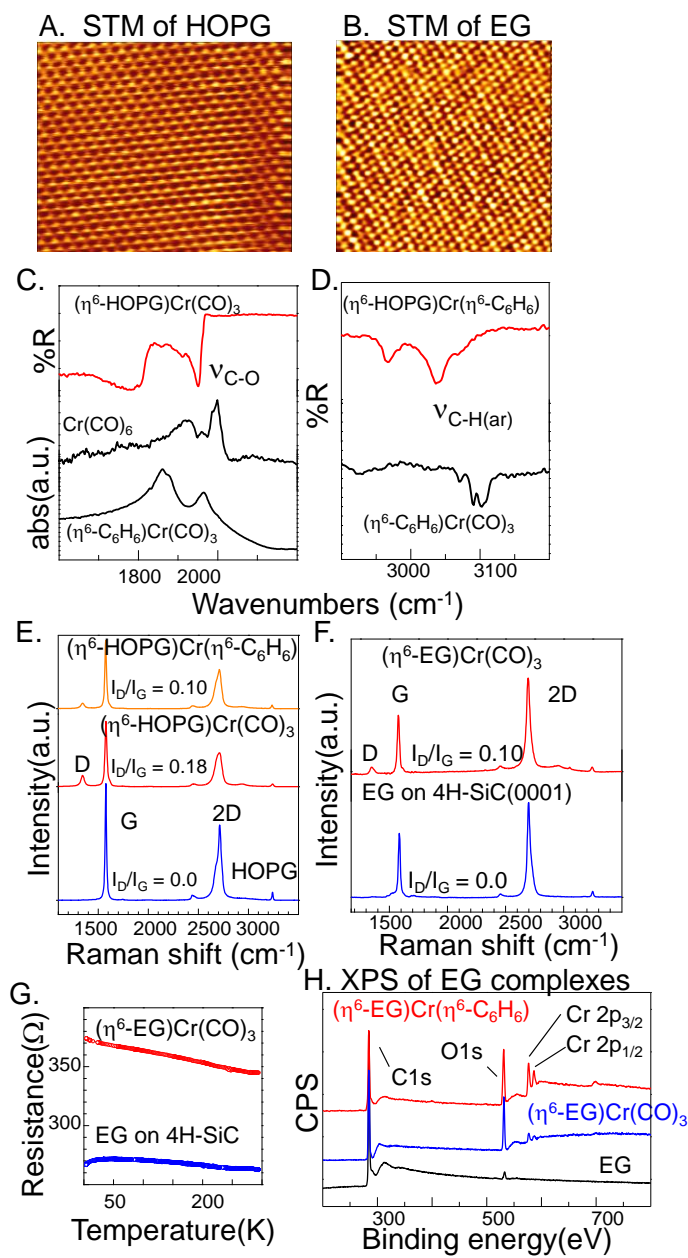

**Fig. 4.** Characteristics of the complexation reactions on HOPG and epitaxial graphene. A) 5 x 3.9 nm STM image of HOPG collected under ambient conditions, $V_s = + 5$ mV, $I_t = 2$ nA. B) 5 x 5 nm STM image of epitaxial graphene, $V_s = + 8$ mV, $I_t = 1$ nA. C) and D) ATR-IR spectra of the chromium reagents and the products with HOPG. E) Raman characteristics of the reaction of $Cr(CO)_6$ and ($\eta^6$-benzene)$Cr(CO)_3$ with HOPG. F) Raman spectra of epitaxial graphene before and after reaction with $Cr(CO)_6$. G) Temperature dependence of sheet resistance of pristine and functionalized epitaxial graphene wafer. H) XPS of pristine EG on 4H-SiC(0001) and the complexation products with the chromium reagents.

## De-complexation of the ($\eta^6$–HOPG)Cr(CO)$_3$ complex: a direct proof of structure

When the ($\eta^6$–HOPG)Cr(CO)$_3$ complex was subjected to an arene exchange reaction (Scheme

2B) in the presence of excess mesitylene, a deep yellow solution was obtained and the Raman spectrum of the HOPG substrate reverted to that of pristine HOPG. The FAB−MS of the yellow colored solution showed the presence of a compound of composition $C_{12}H_{12}O_3Cr$ [m/z: $(M+H)^+=$ 257.0266] (Supporting Information, Fig. S11), indicating that the reaction had led to the dissociation of the HOPG chromium complex with the formation of $(\eta^6-mesitylene)Cr(CO)_3$. The yellow solution in mesitylene turned green on standing, indicating the slow decomposition of the $(\eta^6-mesitylene)Cr(CO)_3$ due to aerial oxidation (similar to that observed for the $(\eta^6-graphene)Cr(\eta^6-benzene)$ complex). This result, together with the shift of the C−O stretching frequency to 1948 $cm^{-1}$ in the FT-IR spectra of the reaction product (compared to the appearance of C−O stretching frequency at 2000 $cm^{-1}$ in $Cr(CO)_6$) (Supporting Information, Fig. S10) confirms the assignment of the reaction product from HOPG and $Cr(CO)_6$, as $(\eta^6-HOPG)Cr(CO)_3$.

**Conclusions**

The organometallic chemistry of graphitic compounds holds great potential for the realization of new materials and we have been able to demonstrate the $\eta^6$−complexation reactions of chromium with various forms of graphene, graphite and carbon nanotubes. These extended periodic π-electron systems exhibit varying degrees of reactivity toward the reagents $Cr(CO)_6$ and $(\eta^6-benzene)Cr(CO)_3$, and we report the formation of $(\eta^6-arene)Cr(CO)_3$ or $(\eta^6-arene)_2Cr$ complexes, where arene = single-walled carbon nanotubes (SWNT), exfoliated graphene (XG), epitaxial graphene (EG) and highly-oriented pyrolytic graphite (HOPG). The compounds were characterized by a combination of IR, Raman, XPS, TGA and further chemical derivatization and we observe clearly understandable trends in the chemistry and stability of the complexes based on curvature and surface presentation. The SWNTs are the least reactive and the least stable, as a result of the effect of curvature on the formation of the hexahapto bond; in the case of substrates with well-defined suraces such as HOPG and EG, we observe the formation of compounds such as $(\eta^6-HOPG)Cr(CO)_3$. The exfoliated graphene (XG) samples were found to give $(\eta^6-graphene)_2Cr$ structures as a result of the accessibility of multiple graphene surfaces to chemical reaction as wells as $(\eta^6-graphene)Cr(CO)_3$ structures depending on the stoichiometry of the chromium reagent used. We report convenient routes for the decomplexation of the

graphene–chromium complexes, which restore the original pristine graphitic surface; exposure of the samples to light or the use of selected ligand competition reactions lead to a reversal of the metal complexation reactions and provide an independent proof of structure for the reaction products. Haptotropic slippage[50] and fluxional behavior[51] of the polyaromatic ligand in ($\eta^6$–arene)–Cr complexes is known to be dependent on the stoichiometry and relative binding affinity of the ligands,[43] which suggests that the chromium (Cr) center will be mobile on graphitic surfaces. This study reports the first use of graphene as a ligand and is expected to expand the scope of graphene chemistry in connection with the application of this material in electronics and organometallic catalysis, where graphene can act as an electronically conjugated catalyst support.

**Acknowledgements:** We acknowledge financial support from DOD/DMEA under contract H94003-10-2-1003 and NSF-MRSEC through contract DMR-0820382. The authors thank Walt de Heer and his group for providing epitaxial graphene samples for this study. The SEM imaging was done at CFAMM-UCR and the MS analysis was done at ACIF-UCR. The XPS measurements were done at UCSB-MRL.

**Notes and references**

*Departments of Chemistry and Chemical and Environmental Engineering, and Center for Nanoscale Science and Engineering, University of California, Riverside, California 92521*

†Electronic Supplementary Information (ESI) available: IR- spectra of the starting chromium reagents and the products, FAB-MS data of the de-complexation reactions, XPS data, absorption spectra, NMR and thermogravimetric analysis data. See DOI: 10.1039/C0SC00634C

# Supporting Information

# Organometallic Complexes of Graphene

Santanu Sarkar, Sandip Niyogi, Elena Bekyarova, Robert C Haddon

*Departments of Chemistry and Chemical and Environmental Engineering, and Center for Nanoscale Science and Engineering, University of California, Riverside, California 92521.*

**Supporting Figures**

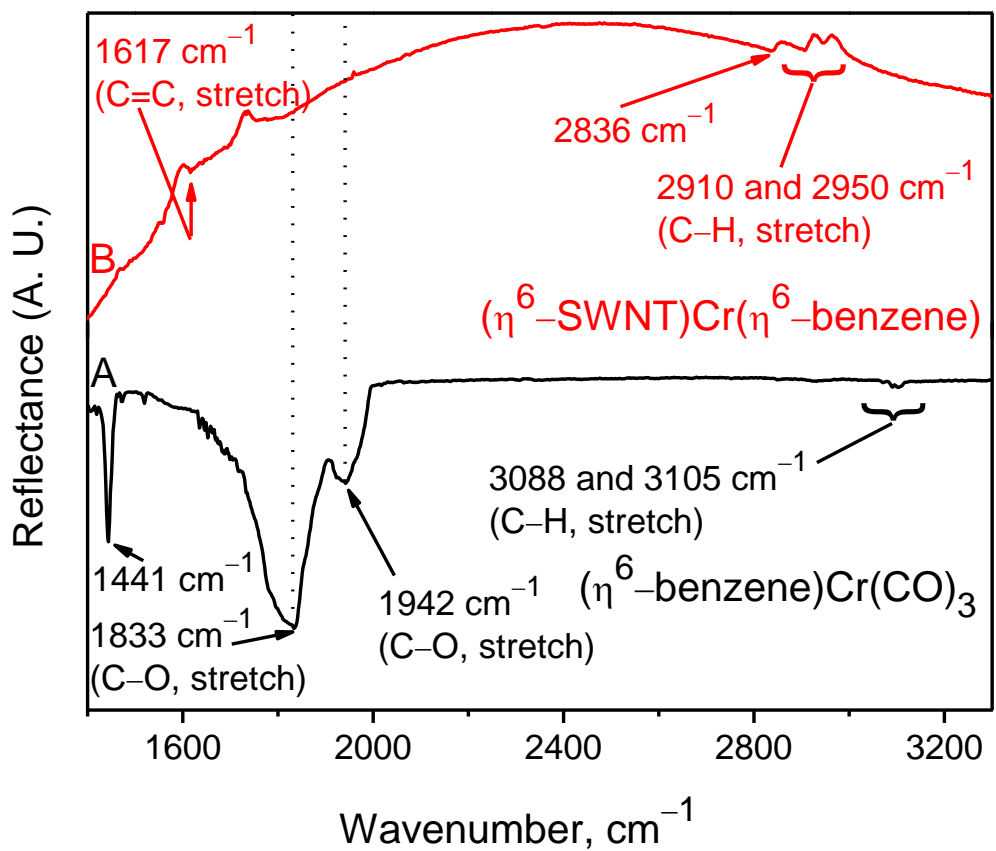

**Figure S1.** The ATR-IR spectra of (A) ($\eta^6$–benzene)Cr(CO)$_3$ and (B) ($\eta^6$–SWNT)Cr($\eta^6$–benzene).

**Table S1.** The C– stretching frequencies of free benzene and various benzene-Cr complexes.

|   | Free ligand and chromium complexes | $C_{sp2}$–H stretches (cm$^{-1}$) |
|---|---|---|
| 1 | Free benzene | 3062[1] |
| 2 | ($\eta^6$–benzene)Cr(CO)$_3$ | 3088, 3105 |
| 3 | ($\eta^6$–benzene)$_2$Cr | 3044, 3080 |
| 4 | ($\eta^6$–SWNT)Cr($\eta^6$–benzene) | 2910, 2950 |

In case of ($\eta^6$–benzene)Cr(CO)$_3$, C$_{sp2}$–H stretches are observed at 3088 and 3105 cm$^{-1}$, which is very close to the stretching frequencies (3090 and 3102 cm$^{-1}$) reported in literature.[2] In case of ($\eta^6$–benzene)$_2$Cr, we observed relatively stronger C$_{sp2}$–H stretches at significantly lower frequency (3044 and 3080 cm$^{-1}$), which is very close to reported values (3058 cm$^{-1}$) in literature.[1] In case of the synthesized ($\eta^6$–SWNT)Cr($\eta^6$–benzene) complex, C$_{sp2}$–H stretches are observed at 2910 and 2950 cm$^{-1}$.

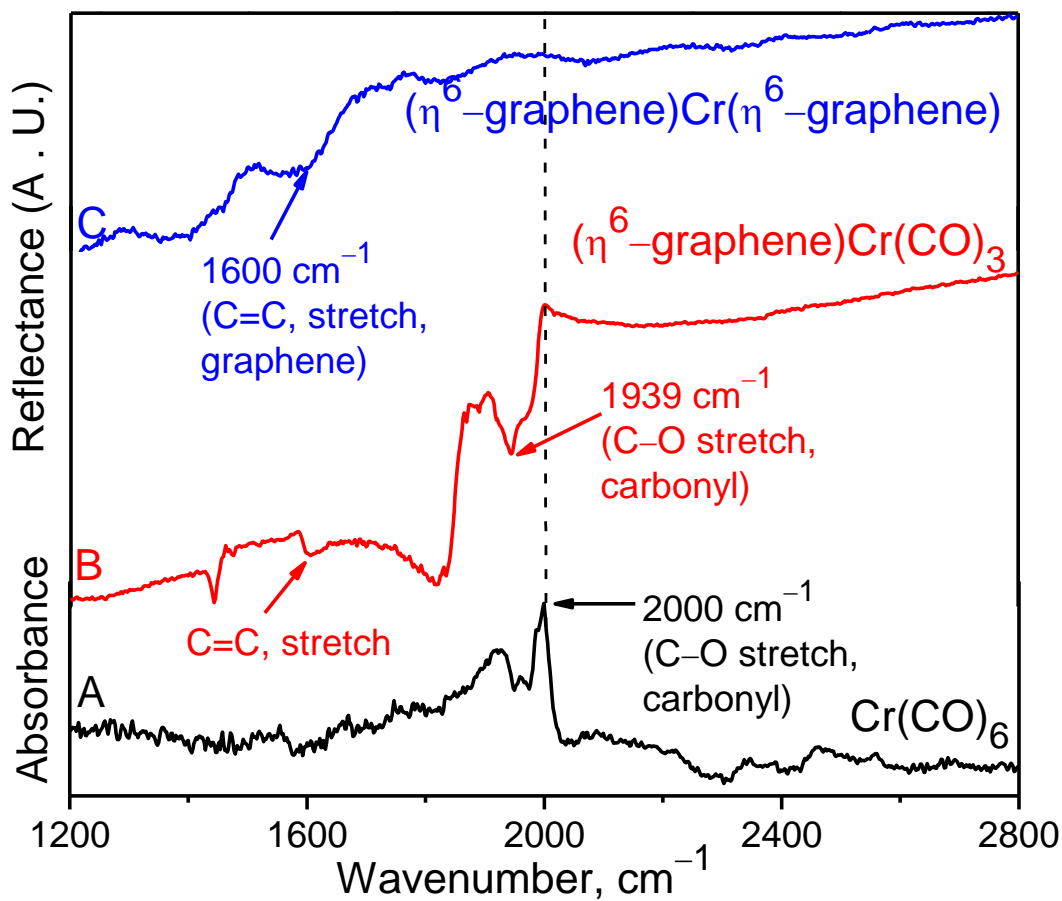

**Figure S2.** The FT-IR spectra (ATR, Ge) of (A) Cr(CO)$_6$, (B) the reaction product between the reaction of XG and Cr(CO)$_6$ (0.13 equivalents), the structure of which is assigned as ($\eta^6$–graphene)Cr (CO)3, and (C) the reaction product between the reaction of XG and Cr(CO)$_6$ (0.02 equivalents), the structure of which is assigned as ($\eta^6$–graphene)Cr($\eta^6$–graphene).

## X-ray Photoelectron Spectroscopy

The XPS measurements were performed at the University of California – Santa Barbara with a Kratos Axis Ultra spectrometer (Kratos Analytical, Manchester, UK) using Al K$\alpha$ monochromated radiation at 1486 eV at base pressure of about $2.10^{-8}$ Torr. The survey spectra were recorded using 270 watts of X-ray power, 80 pass energy, 0.5 eV step size. The high-resolution scans were obtained using power of 300 watts, 40 pass energy and step size of 0.05 eV. A low-energy electron flood from a filament was used for charge neutralization.

Figure S3 illustrates the XPS spectra of graphene (XG)–Cr complexes, where XG is exfoliated graphene, obtained using the solution-exfoliation technique, described in the main text. The Cr 2p spectra of all three chromium complexes – (XG)Cr(CO)$_3$, (XG)Cr(benzene) and (XG)Cr(XG), give a Cr 2p$_{3/2}$ peak at ~577.5 eV, shown in Table 1 along with literature values of molecular chromium complexes. Because of the small chemical shift in the Cr$_{2p}$ lines, the observed peaks cannot be assigned unambiguously to Cr(0) in the complexes of (XG)Cr(CO)$_3$, (XG)Cr(benzene) and (XG)Cr(XG). However, a clear evidence for the presence of CO in the structure of (XG)Cr(CO)$_3$ is observed in the high resolution O1s spectrum, which shows a peak at binding energy of 534 eV.[3] This peak is not present in the other two complexes - (XG)Cr(benzene) and (XG)Cr(XG), indicating absence of the CO ligand in these compounds. The peak centered at 532.5 eV, observed in the spectra of (XG)Cr(benzene) and (XG)Cr(XG), is assigned to oxygen bound to edges and defects in graphene in the form of C=O.[5, 6] We exclude the presence of chromium oxide in these compounds, because O1s appears at ~530 eV in Cr$_2$O$_3$.[7]

**Table S2.** Binding energies (eV) for the ($\eta^6$–graphene)–Cr complexes.

| Complex | Cr(CO)$_6$ | (Bz)Cr(CO)$_3$ | (Bz)$_2$Cr | (XG)$_2$Cr | (XG)Cr(Bz) | (XG)Cr(CO)$_3$ | Cr$_2$O$_3$ |
|---|---|---|---|---|---|---|---|
| O 1s | 534.00[8] | 533.30[3] | - | - | - | 534[3] (due to CO ligand) | 530[7] |
| Cr 2p$_{3/2}$ | 577.6[9] 576.8[3] | 576.1[3] | 575.2[3] | 577.5 | 577.5 | 577.5 | 576.2[4] |

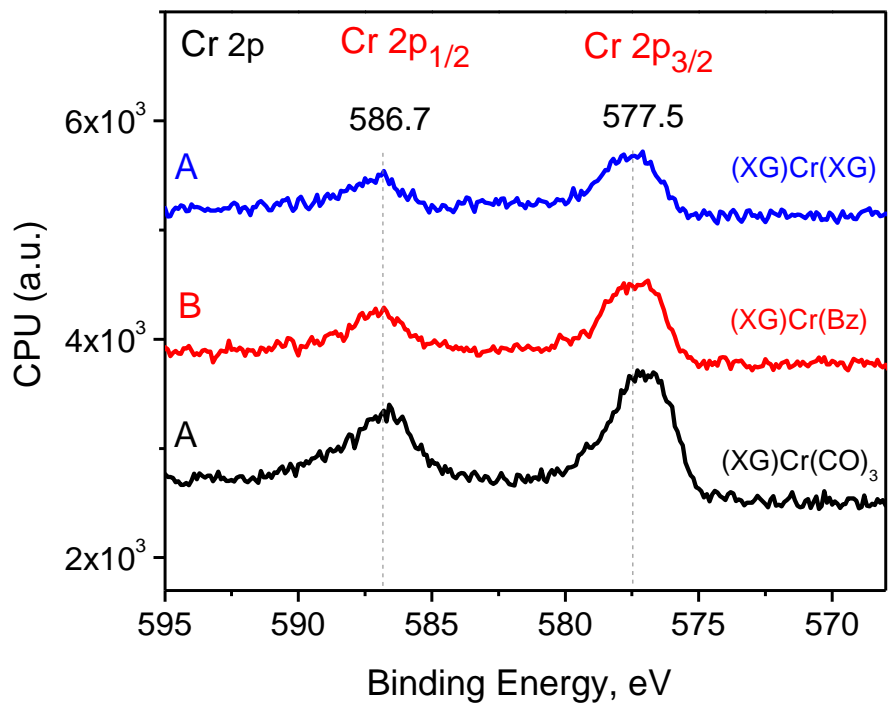
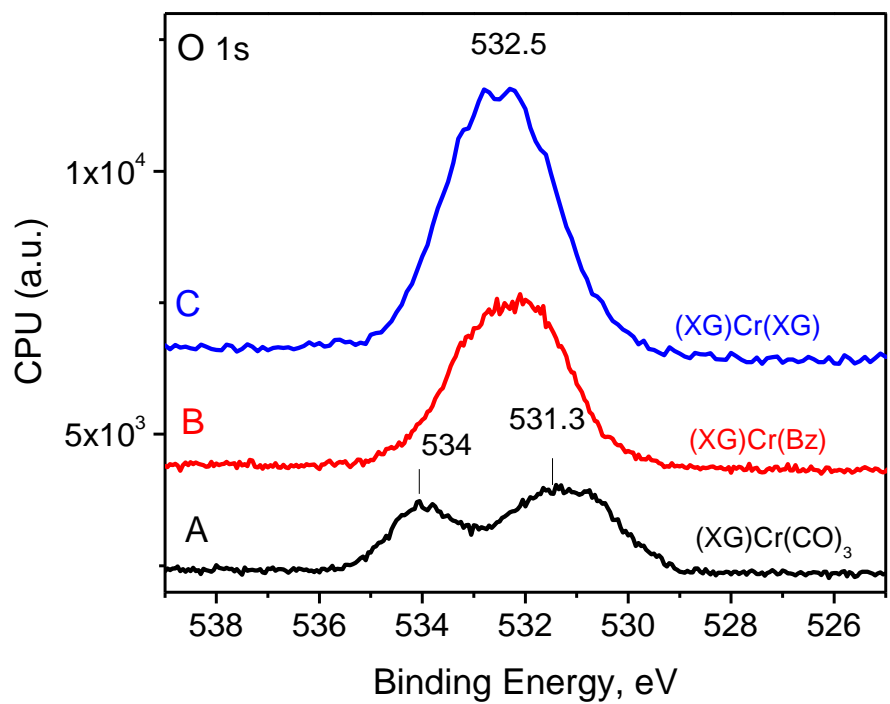

**Figure S3.** XPS data of the synthesized graphene-Cr complexes: (A) (XG)Cr(CO)$_3$, (B) (XG)Cr(benzene) and (C) (XG)Cr(XG), where XG refers to exfoliated graphene.

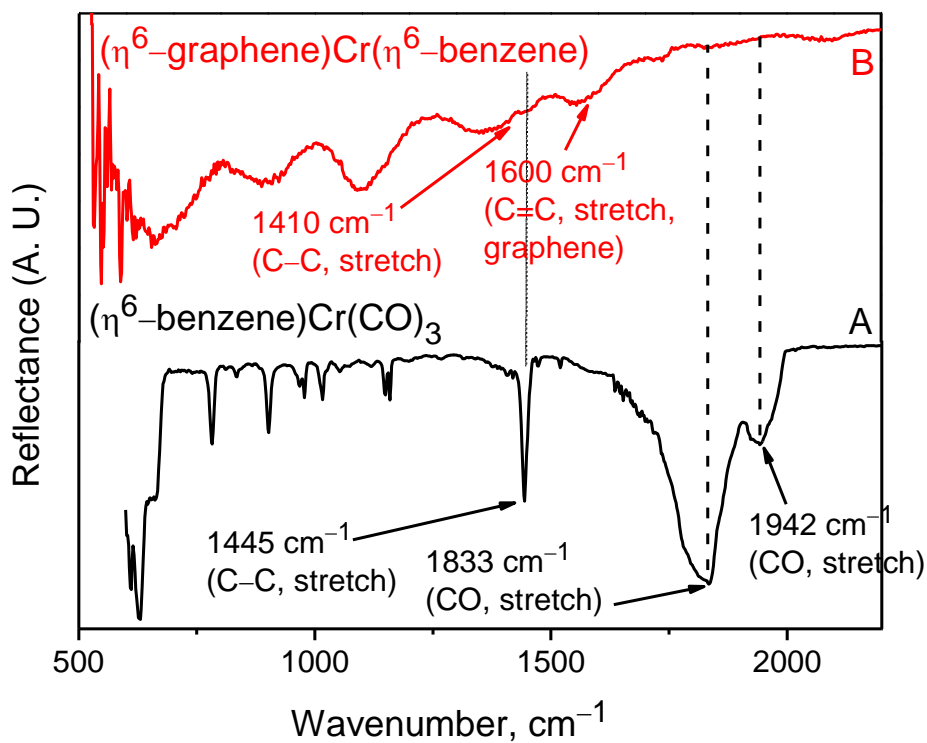

**Figure S4.** The FT-IR spectra (ATR, Ge) of (A) ($\eta^6$–benzene)Cr(CO)$_3$ and (B) ($\eta^6$–graphene)Cr($\eta^6$–benzene).

Assignments of the IR bands of ($\eta^6$–benzene)Cr(CO)$_3$ can be found literature[2] and the important bands are reported to appear at 535, 575, 636, 653, 669, 1444 (C–C stretch), 1854 (C–O stretch, E), 1954 (C–O stretch, A$_1$), 3070 and 3102 (C–H stretch) cm$^{-1}$. Thus in the case of reaction between XG and ($\eta^6$–benzene)Cr(CO)$_3$, the absence of C–O stretching vibrations in the reaction product supports the assignment of the reaction product as ($\eta^6$–graphene)Cr($\eta^6$–benzene). This is further confirmed after $^1$H-NMR analysis of the de-complexation products of the resulting graphene-Cr complex with mesitylene, as discussed later.

**Discussion of TGA Analysis**

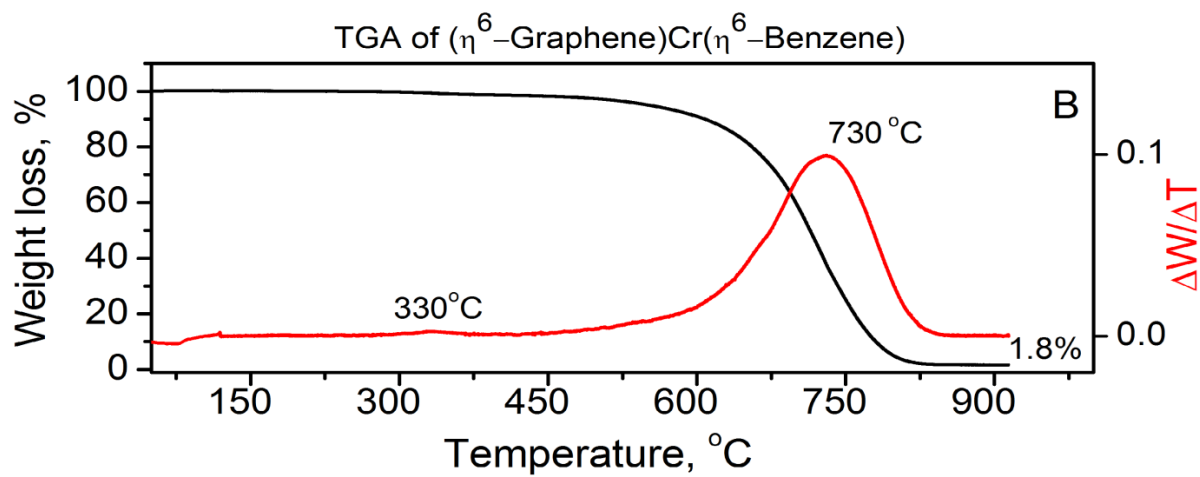

**Figure S5.** Thermogravimetric analysis of ($\eta^6$–graphene)Cr($\eta^6$–benzene) in air, showing the 1.8 wt% residue which is obtained in the form of a dark green powder.

The green-colored residue obtained after the TGA experiment in air is ascribed to chromium oxide ($Cr_2O_3$), although a small fraction chromium carbide ($Cr_3C_2$) could also be formed at this temperature range. Chromium carbides are generally formed by reactions between metallic chromium and carbon (graphite) under oxygen-free inert (argon) atmosphere at 800 °C.[10]

**Decomplexation reactions with Mesitylene**

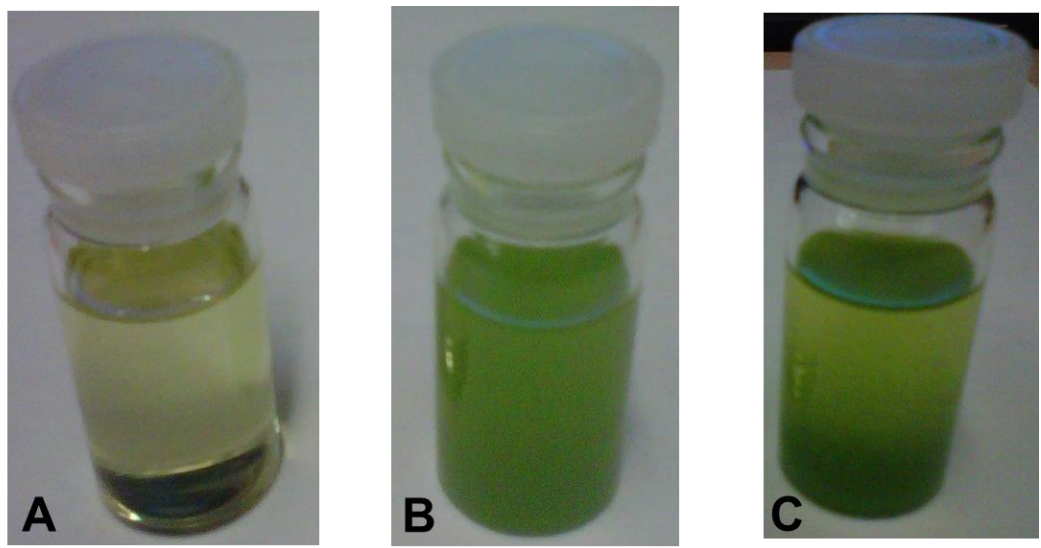

**Figure S6.** Representative time lapse photographs of solutions of the mesitylene−chromium complexes obtained after de-complexation reactions of mesitylene with ($\eta^6$−graphene)Cr($\eta^6$−benzene). (A) Yellow solution of the mesitylene complex, (B) aerial decomposition of the solution upon standing for a few hours in air to give a green colored solution, and (C) precipitation of chromium as a green solid upon standing overnight in air.

The de-complexation reactions of ($\eta^6$−graphene)Cr($\eta^6$−graphene) with mesitylene was also performed similarly by heating the complex at 150 °C under argon atmosphere overnight; the solid was collected upon filtration. The yellow filtrate of mesitylene-Cr complexes obtained during the filtration immediately precipitated as a green solid with a colorless supernatant. The mesitylene-Cr complexes are known to be very unstable and their structural chemistry is complicated due to formation of multidecker complexes of various nuclearities.[11, 12] Therefore, the FAB-MS analysis of the filtrate often shows mixture of compounds of higher masses with m/z values ranging from 315.2581 to 663.4486, which were difficult to assign.

# $^1$H NMR (CDCl$_3$) spectroscopy of the yellow mesitylene solution

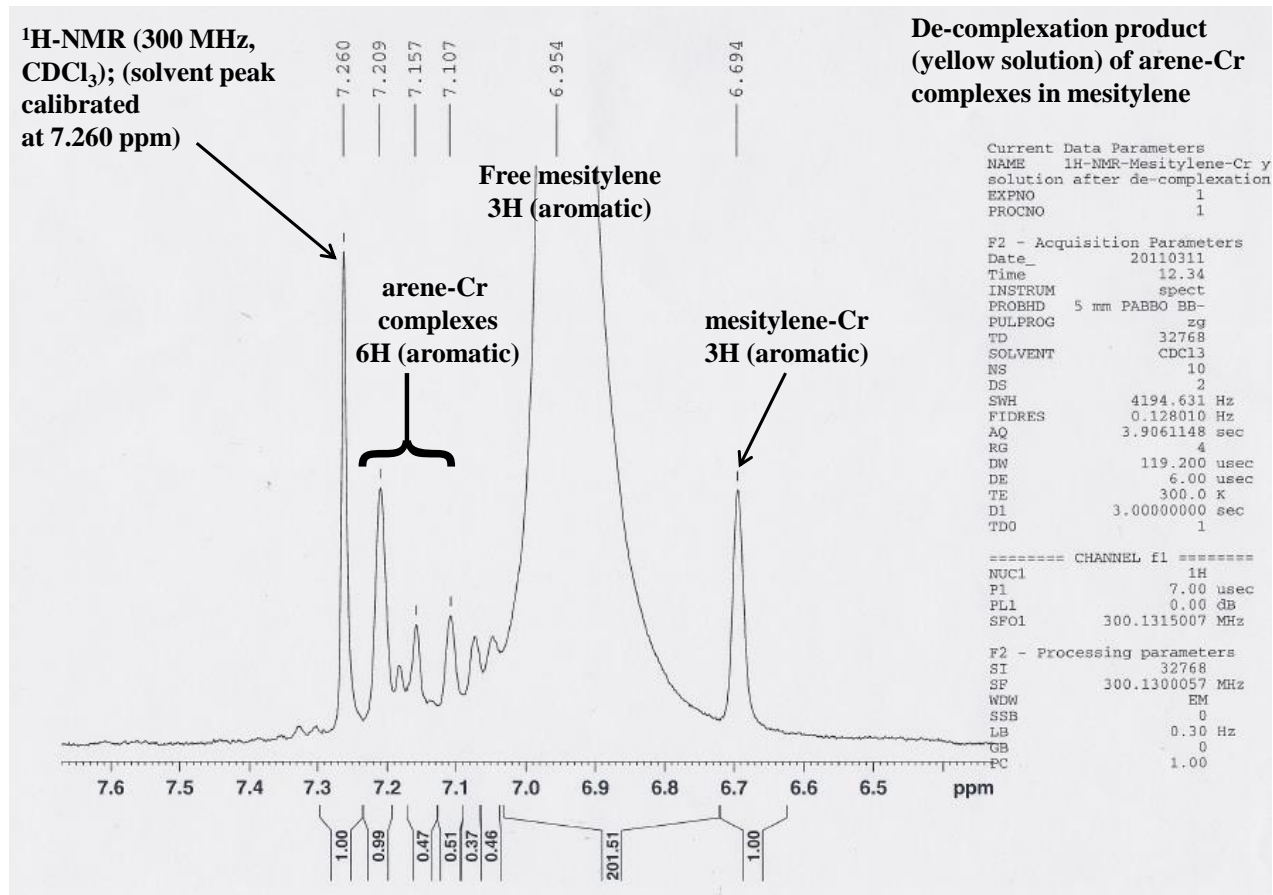

**Figure S7.** $^1$H-NMR (300 MHz, CDCl$_3$) spectra of the yellow solution obtained after de-complexation of ($\eta^6$–graphene)Cr($\eta^6$–benzene) with excess mesitylene showing the presence of ($\eta^6$–mesitylene)Cr($\eta^6$–benzene) complexes in excess free mesitylene. The aromatic region of the proton NMR spectra of the yellow mesitylene solution of mesitylene-chromium complex in CDCl$_3$ is expanded.

$^1$H-NMR spectra (in CDCl$_3$) of free benzene, ($\eta^6$–benzene)Cr(CO)$_3$, ($\eta^6$–benzene)$_2$Cr shows resonances at 7.36,[13] 5.30 (5.26 ppm)[13] and 7.16 ppm respectively, while in case of the product obtained after de-complexation of ($\eta^6$–graphene)Cr($\eta^6$–benzene) with excess mesitylene shows the presence of resonances at 7.209, 7.157 and 7.107 ppm (total integration corresponds to 6H), which could be tentatively assigned to six aromatic protons of the benzene ring, and at 6.694 ppm (3H, aromatic) which could be assigned

as the three aromatic protons of the mesitylene ring; consequently the yellow mesitylene shows a signature of the formation of ($\eta^6$–mesitylene)Cr($\eta^6$–benzene) complex after de-complexation reaction. The presence of excess free mesitylene (used for de-complexation) is attested by the presence of resonances at 6.954 ppm, which are due to aromatic protons (3H) of mesitylene.

**Ultraviolet spectra of chromium complexes**

The UV spectra of the reagents – $Cr(CO)_6$ and ($\eta^6$–benzene)Cr(CO)$_3$, and the yellow product [($\eta^6$–mesitylene)Cr($\eta^6$–benzene)] obtained after de-complexation of ($\eta^6$–graphene)Cr($\eta^6$–benzene) with excess mesitylene are shown in Figure S8. The spectra were recorded on a Cary 5000 spectrophotometer using ethanol as a solvent. The spectra of $Cr(CO)_6$ and ($\eta^6$–benzene)Cr(CO)$_3$ show absorption maxima at 279 and 312 nm, respectively. This electronic absorption bands have been attributed to be typical for transition metal-carbon bonds.[14] Spectra of a concentrated solution of the yellow product also showed a broad peak at wavelength of ~315 nm (Figure S8B). Upon dilution of the solutions this peak disappeared and the spectra revealed a high-intensity maximum at 205 nm with a shoulder at ~215 nm. The spectrum of the diluted solution of the product of the de-complexation reaction is almost identical to that of pure mesitylene and we assign the disappearance of the peak at 315 nm to the fact that the concentration of the product falls below the spectrometer's detection limit. The observed peaks in the low wavelength range are due to the excess of mesitylene in the reaction mixture. The observed peak at 315 nm confirms the formation of

($\eta^6$−mesitylene)Cr($\eta^6$−benzene) obtained after de-complexation of

($\eta^6$−graphene)Cr($\eta^6$−benzene) with excess mesitylene.

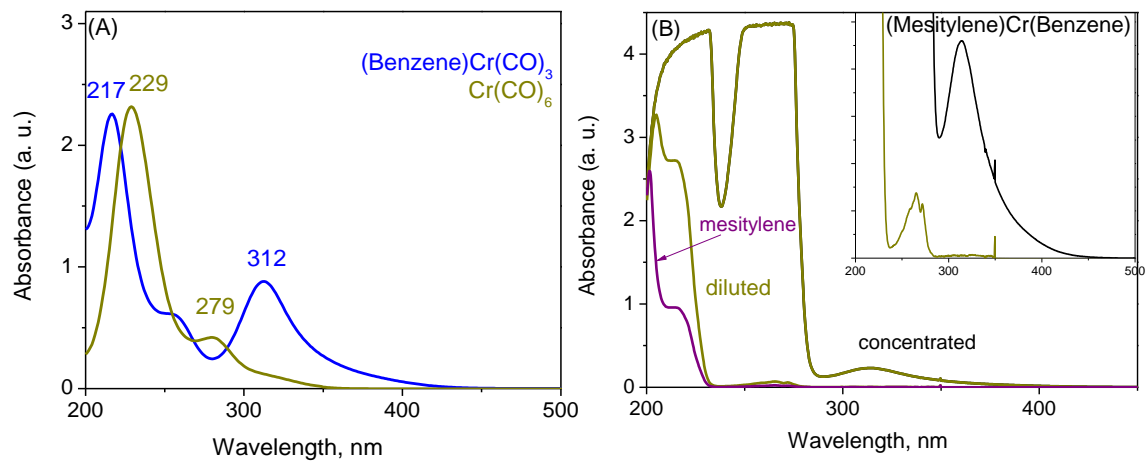

**Figure S8.** Optical absorption spectra of the molecular chromium complexes and of the yellow solution obtained after de-complexation of ($\eta^6$−graphene)Cr($\eta^6$−benzene) with excess mesitylene.

## Mass spectroscopic analysis of de-complexation reaction of ($\eta^6$–graphene)Cr($\eta^6$–benzene) with mesitylene

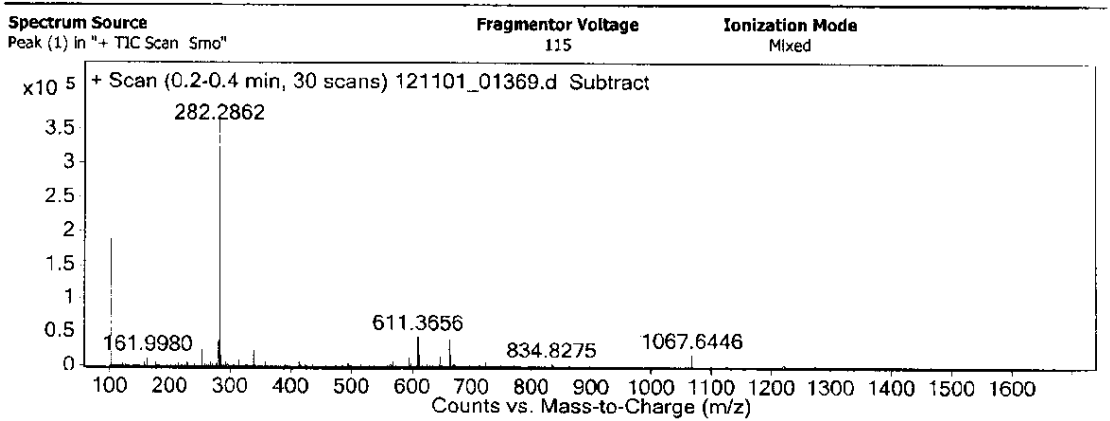

**Figure S9.** FAB-MS analysis of the mesitylene–chromium complex (Figure S6) obtained from de-complexation of ($\eta^6$–graphene)Cr($\eta^6$–benzene) with mesitylene.

Analysis of the yellow solution (in Figure S6) obtained after the de-complexation reaction of ($\eta^6$–graphene)Cr($\eta^6$–benzene) with excess mesitylene by FAB-MS technique shows a composition with a parent mass of m/z = 282.2862, which could neither be assigned to ($\eta^6$–mesitylene)Cr($\eta^6$–benzene) (molar mass = 250.30) nor ($\eta^6$–mesitylene)$_2$Cr (molar mass = 292.38). These arene complexes are known to be labile.[11] However, based on the changes in chemical shifts and proton integration values in $^1$H-NMR spectra (Figure S7) the structure of the complex was tentatively assigned to ($\eta^6$–mesitylene)Cr($\eta^6$–benzene).

S12

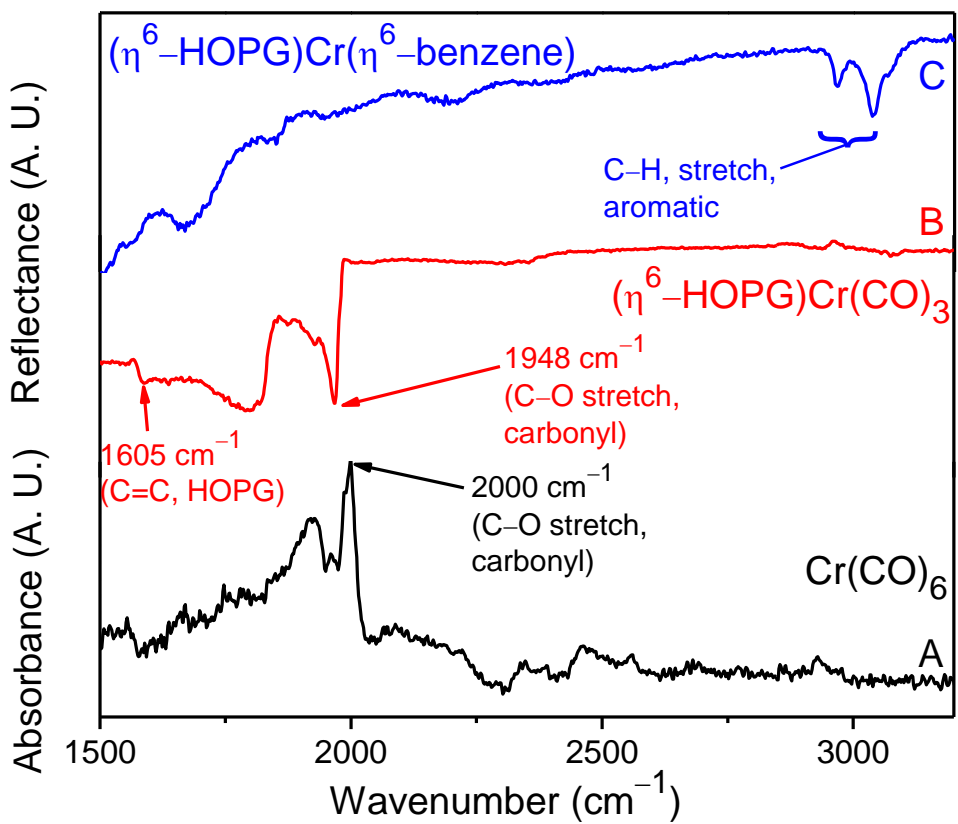

**Figure S10.** The FT-IR spectra (ATR, Ge) of (A) $Cr(CO)_6$, (B) $(\eta^6-HOPG)Cr(CO)_3$ and (C) $(\eta^6-HOPG)Cr(\eta^6-benzene)$.

The IR bands of $Cr(CO)_6$ are reported[15] to appear at 2000 (C–O stretch due to carbonyl ligands), 1965 (combination band) and 668 (Cr–C stretching vibration) $cm^{-1}$. In the present case the complexation of HOPG with the chromium tricarbonyl moiety results in a shift in the C–O stretching frequency of the carbonyls from 2000 ($Cr(CO)_6$) to 1948 $cm^{-1}$ in the product, indicating the successful grafting of the $-Cr(CO)_3$ moiety to the top layer of the HOPG surface.

The substitution of three CO ligands in $Cr(CO)_6$ with the electron-rich graphene ligand should increase the extent of the metal $d_\pi$ back-donation[16] to the remaining CO groups;

consequently the Cr–CO bond order is expected to increase and the C–O bond order of the residual carbonyl ligands is expected to decrease. This is reflected in the spectra obtained after coordination of HOPG where it may be seen that the C–O stretching frequency is 50 cm$^{-1}$ lower in ($\eta^6$–HOPG)Cr(CO)$_3$ than in Cr(CO)$_6$, and is comparable to the C–O stretching frequencies of ($\eta^6$–benzene)Cr(CO)$_3$ [1854 (C–O stretch, E) and 1954 (C–O stretch, A$_1$) cm$^{-1}$].[2]

The absence of similar C–O stretching vibrations in the reaction product between HOPG and ($\eta^6$–benzene)Cr(CO)$_3$ provides the basis for formulation of the reaction products as ($\eta^6$–HOPG)Cr($\eta^6$–benzene). The case is supported by the thermochemical study on the mean bond dissociation energy of the Cr–arene bond by Connor et al.[17]

## Mass spectroscopic analysis of de-complexation reaction of ($\eta^6$–HOPG)Cr(CO)$_3$ with mesitylene

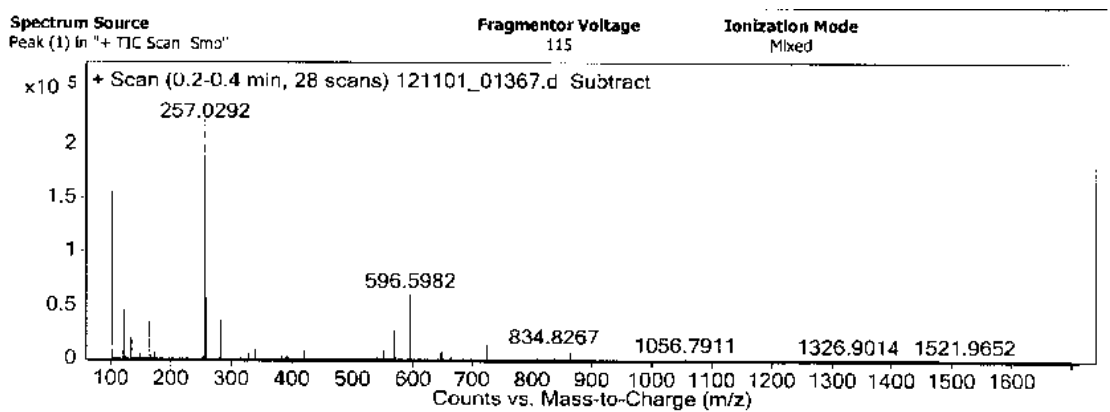

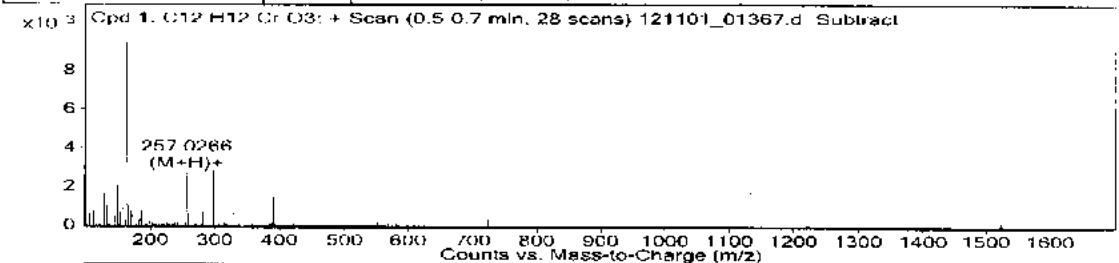

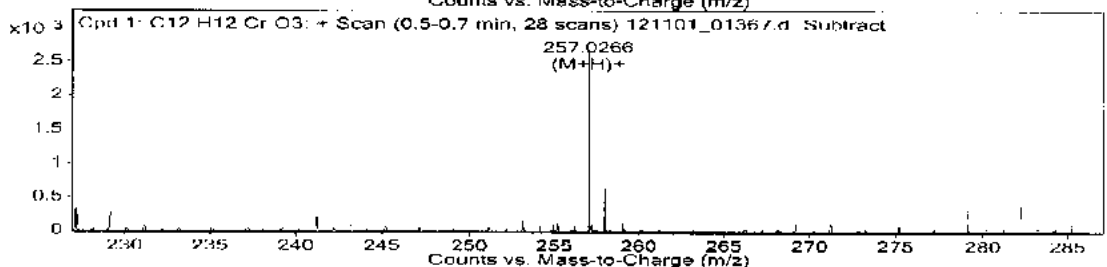

**Figure S11.** FAB-MS analysis of the mesitylene–chromium complexes obtained after de-complexation of the reaction product of HOPG and Cr(CO)$_6$.

When the product of the reaction of HOPG and $Cr(CO)_6$ is de-complexed via arene exchange reactions using excess mesitylene, it results in a deep yellow solution which shows the presence of a compound of composition $[(\eta^6-mesitylene)Cr(CO)_3H]^+$ with m/z = 257.0266 on analysis with FAB-MS. This result along with observed shift in the IR stretching frequency of the remaining CO ligands (Figure S7B) provides persuasive evidence that that the hexahapto-complexation of HOPG with $Cr(CO)_6$ results in the formation of $(\eta^6-HOPG)Cr(CO)_3$, which when subjected to an arene exchange reaction with mesitylene, dissociates to give a metal-free graphene HOPG surface and $(\eta^6-mesitylene)Cr(CO)_3$ (molar weight of 256.22), which forms a deep yellow solution in mesitylene.

## Supporting References